\documentclass[lettersize,journal]{IEEEtran}
\usepackage{amsmath,amsfonts}
\usepackage{algorithmic}
\usepackage{array}
\usepackage[caption=false,font=normalsize,labelfont=sf,textfont=sf]{subfig}
\usepackage{stfloats}
\usepackage{verbatim}
\usepackage{graphicx}
\usepackage{tikz}
\newcommand*\numrounded[1]{\tikz[baseline=(char.base)]{
            \node[shape=circle,draw,inner sep=0.7pt] (char) {#1};}}
\usepackage{acro}
\usepackage{booktabs}
\usepackage{color}
\usepackage{tabularx}
\usepackage{hyperref} 
\hypersetup{
    colorlinks=true,
    linkcolor=blue,
    filecolor=magenta,      
    urlcolor=cyan,
}
\hypersetup{linkcolor=black}
\hypersetup{citecolor=black}
\begin{document}

\title{Large Language Models for Zero Touch Network Configuration Management}

\author{\IEEEauthorblockN{Oscar G. Lira\IEEEauthorrefmark{1}, Oscar M. Caicedo\IEEEauthorrefmark{2}, Nelson L. S. da Fonseca\IEEEauthorrefmark{1}}
\thanks{O. G. Lira and N.L.S. da Fonseca are with the Institute of Computing, State University of Campinas, Campinas 13083-852, Brazil. E-mail: \{o224415,nfonseca\}@ic.unicamp.br.}
\thanks{O. M. Caicedo is with the Departamento de Telem\'atica, Universidad del Cauca, Popay\'an, 19001, Colombia. E-mail: omcaicedo@unicauca.edu.co.}
}

\markboth{IEEE Communications Magazine, ~Vol.~XX, No.~XX, May~2024}%
{Shell \MakeLowercase{\textit{et al.}}: A Sample Article Using IEEEtran.cls for IEEE Journals}

\maketitle

\begin{abstract}
The Zero-touch Network \& Service Management (ZSM) paradigm, a direct response to the increasing complexity of communication networks, is a problem-solving approach. In this paper, taking advantage of recent advances in generative Artificial Intelligence, we introduce the   \textbf{Net}work \textbf{C}on\textbf{F}iguration \textbf{G}enerator (LLM-NetCFG) that employs \textbf{L}arge \textbf{L}anguage \textbf{M}odel and  architects ZSM configuration agents by Large Language Models. LLM-NetCFG can automatically generate configurations, verify them, and configure network devices based on intents expressed in natural language. We also show the automation and verification of network configurations with minimum human intervention. Moreover, we explore the opportunities and challenges of integrating LLM in functional areas of network management to fully achieve ZSM.
\end{abstract}


\section{Introduction}
The European Telecommunications Standards Institute (ETSI) defined a network management paradigm to realize network self-operation, self-maintenance, and self-optimization with minimal (or without) human intervention, named  Zero-touch Network \& Service Management (ZSM) \cite{ETSI-ZSM}. Self-configuration, the ability to configure network parameters without human intervention,  is an essential building block for ZSM. In addition, the Resource Configuration Management Service (RCMS)  is a crucial component of ZSM for coordinating network configuration, improving operational agility, and reducing policy errors. RCMS oversees the network configuration management across network domains, empowering the network with autonomous configuration generation and deployment.

Intent-driven networks (IDN) \cite{10032317} allow operators to describe their business objectives and requirements using natural language to reduce the complexity of network administration. ETSI defined intents to enhance and assist the RCMS by adopting actionable directives and translating them into policies and precise configuration parameters. IDN operation relies on effectively resolving intents to achieve operators' goals. 
Translating and solving intents requires advanced language processing and network knowledge.

A large language model (LLM) can create human-like text by processing large amounts of data. LLMs can be integrated
into IDN networks to achieve ZSM, especially to generate
network and device configuration recommendations. In this regard, \cite{ref24} proposed NETBUDDY, a system that aims to simplify and automate the configuration of network devices using GPT-4. The authors in \cite{ref25} explored LLMs to create recommendations on network device configuration using instruction prompts in GPT-4. However, these models demand large amounts of computational resources due to their size and consequently many accesses to remote platforms. Such an approach implies sharing sensitive information over the Internet, increasing the vulnerability to information leakage and unauthorized access to devices, even when sent over secure channels. 

This paper introduces the  \textbf{L}arge \textbf{L}anguage \textbf{M}odel-based \textbf{Net}work \textbf{C}on\textbf{F}iguration \textbf{G}enerator (LLM-NetCFG) to address the threat of disclosing network configuration information. LLM-NetCFG employs a local LLM and avoids using an Internet-accessible LLM platform to translate high-level intents into specific network device configurations by locally executing the Zephyr-7b beta model. This approach ensures the privacy of information about network devices and infrastructure. Additional original contributions in this paper are: \numrounded{1} a showcase of leveraging a local LLM for configuring  Zero-touch networks; \numrounded{2} a showcase of verification of network configurations with minimum human intervention in a local environment; \numrounded{3} a discussion of research opportunities and challenges to build LLM-based ZSM agents.

This paper is organized as follows. Section II reviews Large Language Models. Section III discusses the application of LLMs in network management. Section IV covers LLMs in ZSM. Section V presents the operation of LLM-NetCFG. Section VI presents a use case. Section VII discusses future research. Finally, Section VIII concludes the paper.

\section{Large Language Models}

\subsection{Architectures}
LLMs are AI applications with Natural Language Processing (NLP) capabilities, which allow machines to understand and mimic human language. Those capabilities come from using deep neural networks with a self-attention mechanism that enables the model to dynamically weigh the relevance of each word to contextualize and understand the input text meaning word by word.

A transformer is a deep neural network (DNN) architecture (model) with attention mechanisms. Transformers efficiently enable self-attention mechanisms by processing multiple input sequences simultaneously, speeding up training and inference processes. Encoder-only transformers use self-attention mechanisms to capture dependencies between different words regardless of their position by considering only the input data. They produce a meaningful, fixed-size output representation that can be used for various downstream tasks (e.g., classification tasks). Decoder-only transformers generate coherent text by predicting the following words in a sequence, considering previously generated words using the self-attention mechanisms (e.g., summarization and translation tasks). Encoder-decoder transformers combine the encoder to process the input and the decoder to generate the output. 

LLMs are versatile in handling various NLP tasks. However, their performance is influenced by factors such as the model's comprehension of inputs, the efficacy of its training process, and the expertise with which users interact to accomplish tasks' objectives. The combination of data preprocessing (i.e., tokenization), structured learning (i.e., training and fine-tuning), and strategic guidance (i.e., prompts and prompt engineering) ensures the customization of the model to perform specialized tasks with high accuracy and efficiency.

\subsection{Tokenization and Training}
Tokenization is a crucial step in LLMs forming the foundation for data understanding. Tokens, the discrete parts of a text, are the building blocks that LLMs use to comprehend the data.
Tokenization involves separating a text into discrete parts. Training enables LLMs to comprehend and generate human-like text with coherence by learning the probability distribution of the vocabulary from a vast corpus using diverse techniques such as next-token prediction, masked language, n-grams, and continuous space. 

\subsection{Prompts and Prompt engineering}
A prompt is a statement or question that guides an LLM output. It sets the context and specifies the task required by a system, an assistant, or a user role. In the system role, a well-designed prompt ensures that the LLM understands the user's request accurately to generate a relevant response. The assistant role carries out a desired task based on the user's input, understands the context of the user's request, and tailors the response accordingly. The user role encompasses individual interactions with the LLM by inputting queries, instructions, or commands to get responses or perform tasks. 

Prompt engineering determines the relevance (applicability and usefulness) and quality (accuracy and time metrics) of the responses given by LLM-based models. An effective prompt not only comprises carefully chosen words, but also emphasizes their arrangement, inclusion of specific keywords or phrases, or even the incorporation of example responses. Priming, zero-shot, one-shot, few-shot, chain-of-thought, and self-consistency are prompt engineering techniques. 

\subsection{Fine-tuning}
Fine-tuning involves training a pre-trained model on a specific task or dataset, seeking to refine the model's general knowledge and adapt it to the particularities of the new task. The adaptation facilitated by fine-tuning and the capacity for language understanding and handling massive amounts of knowledge has positioned LLMs as effective assistants for humans and promoted their incorporation into complex and diverse domains.
Transfer learning and reinforcement learning can be used in fine-tuning techniques. 

\section{LLMs for Network Management Functionalities: What Has Been Done?} 
In the literature, researchers have recently started constructing domain-adapted LLMs for the networking field \cite{huang2023l} and LLM-based network management functional areas: fault, configuration, accounting, performance, and security (FCAPS).

\textbf{Fault:} NetLM \cite{ref15} introduces a network generative AI-based architecture designed to understand the sequence structures in the network data packets for 6G networks, predict maintenance, and perform troubleshooting tasks. RCAgent \cite{wang2023rcagent} proposes a practical and privacy-aware industrial root cause analysis solution. It explores LLMs for building customized conversational interfaces in cloud environments.

\textbf{Configuration:} NETBUDDY \cite{ref24} addresses the traditionally complex and error-prone task of configuring network devices by using GPT-4 to generate human-readable configurations and actionable code for border routers and programmable switches. \cite{ref25} introduces pair programming and verified prompt programming as methodologies to achieve GPT-4-human collaboration to synthesize correct router configurations crafted by network administrators, translate configurations from one network device brand to another one, and verify such configurations with minimal human intervention.

\textbf{Accounting:} \cite{soman2023observations} evaluates the usage of LLMs in conversational interfaces within the telecommunications domain. The experiments were applied to various chat interfaces with questions related to 4G and 5G networks and device and resource management. The findings demonstrate that these models exhibit strong baseline performance and highlight the need for fine-tuning LLMs on domain-specific tasks to ensure reliable enterprise-grade applications.

\textbf{Performance:} \cite{10384606} introduces a framework based on Generative Pretrained Transformers (GPTs), which automatically organizes, adapts, and optimizes performance in edge systems.  \cite{zhang2024interactive} investigates how to improve the user experience and promote efficient network management with an AI-enabled network management framework consisting of environment, perception, action, and brain units that make decisions considering a knowledge base built with
an LLM module and a Retrieval Augmented Generation module. NetGPT \cite{chen2023netgpt} proposes on-device LLMs for multi-agent intelligence in wireless zero-touch networks to improve user experience and network performance.

\textbf{Security:} SecureBERT \cite{aghaei2022securebert} is a cybersecurity solution that leverages LLMs to detect and prevent fraud efficiently. This solution was trained on a vast corpus of customer data and security-related materials, including articles, books, surveys, blogs, and reports on threats, attacks, and vulnerabilities.

Using LLMs to realize FCAPS advances network management. However, ZSM requires the application of LLMs throughout its entire architecture to ensure end-to-end automation and orchestration. ZSM emphasizes comprehensive, autonomous network operations, which requires integrating LLMs in all layers and processes within the network to fully realize zero-touch management and service delivery.

\section{LLM-aided ZSM} 
To achieve ZSM objectives, supervised, unsupervised, semi-supervised, deep, and reinforcement learning models have been applied. However, to our knowledge, the use of LLM in ZSM has not yet been fully explored. LLMs can create conversational interfaces for network management, allowing operators to query the system using natural language and receive comprehensible explanations of AI-driven decisions with low complexity. Furthermore, LLMs can improve the automation of performance management and root cause analysis by providing more sophisticated analytical capabilities and improving the system's ability to learn from data, thus contributing to ZSM objectives such as self-management capabilities, closed-loop automation, and intent-driven management \cite{LIYANAGE2022103362}.

\begin{figure}[!hb]
    \centering
    \caption{LLMs Into ZSM Components.}
    \includegraphics[width=0.49\textwidth]{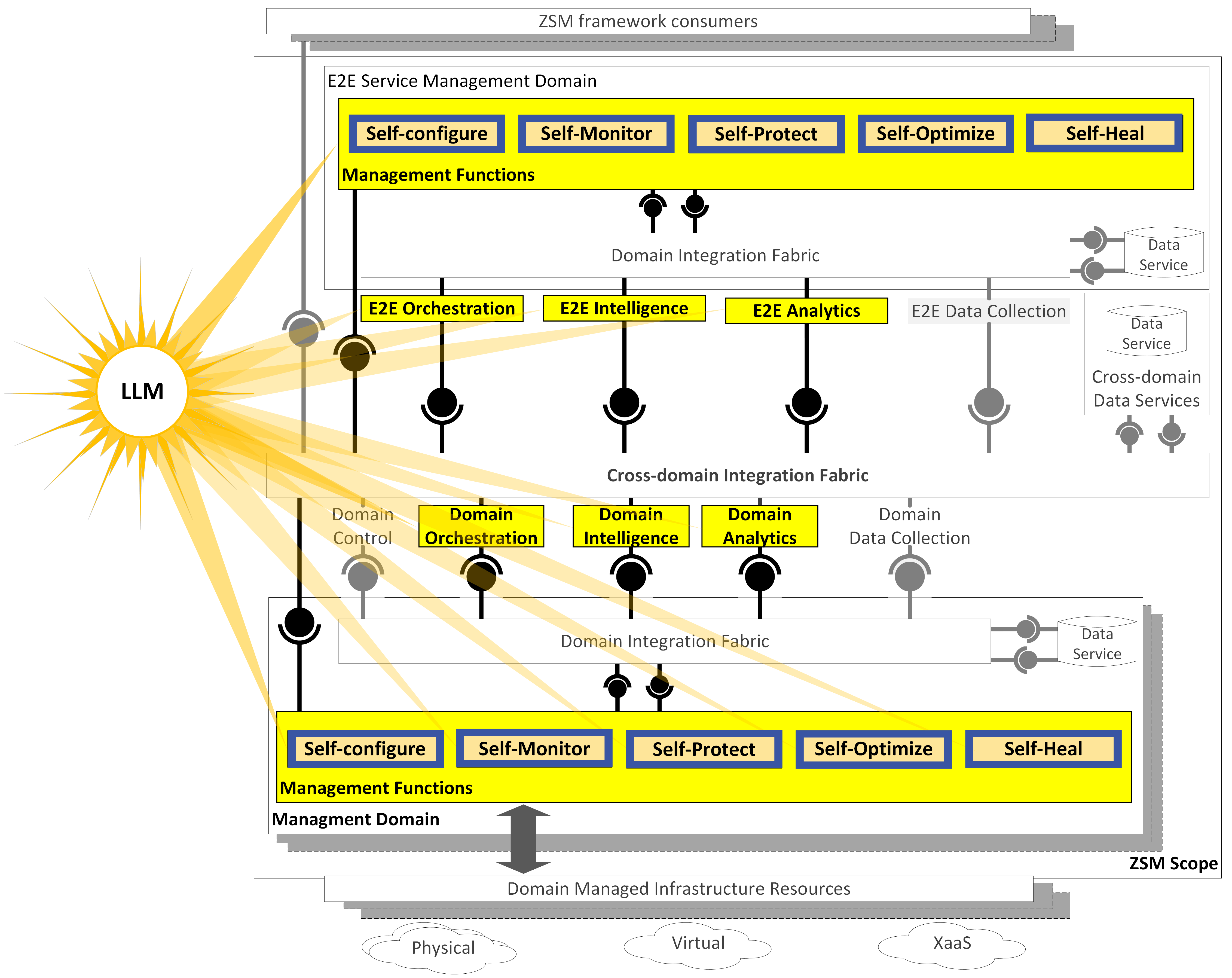}
    \label{fig_Ilum}
\end{figure}

ZSM can be deployed as an intent-driven autonomous network that can perform self-configuration, self-monitoring, self-healing, self-optimization, and self-protection without human intervention. Figure~\ref{fig_Ilum} shows the ZSM architecture that highlights the components in which we consider LLMs to help manage the zero-touch network. The ETSI ZSM architecture comprises various components. \textit{E2E orchestration} coordinates resources and services across multiple management domains. \textit{E2E intelligence} leverages data analytics and AI for predictive analytics, anomaly detection, optimization, and adaptive learning. \textit{Domain analytics} and \textit{E2E analytics} are crucial to gaining insights by analyzing data within a domain or throughout the network. \textit{E2E service management} ensures that services are delivered, maintained, and optimized by interacting with the \textit{orchestration} and \textit{intelligence} components. \textit{Management functions} are responsible for performing autonomous tasks.  \textit{Domain data collection} and \textit{E2E data collection} focus on gathering data to inform local or cross-domain management activities. The \textit{Integration Fabric} facilitates data exchange between these domains, ensuring seamless connectivity and a unified data environment. \textit{Cross-domain Data Services} enable data storage, sharing, and analysis across these domains, further supporting a unified data environment. \textit{Domain control} manages and enforces policies within each domain, ensuring  they align with overall \textit{orchestration} and \textit{intelligence} strategies. Closed-loop architecture automation, which includes observing, orienting, deciding, and acting functions, uses \textit{ cross-domain data services} to support self-configuration, self-monitoring, self-healing, self-optimization, and self-protection capabilities, facilitating a fully autonomous network management system.

\begin{figure*}[h!]
\centering
\caption{LLM-NetCFG Structure.}
\includegraphics[width=\textwidth]{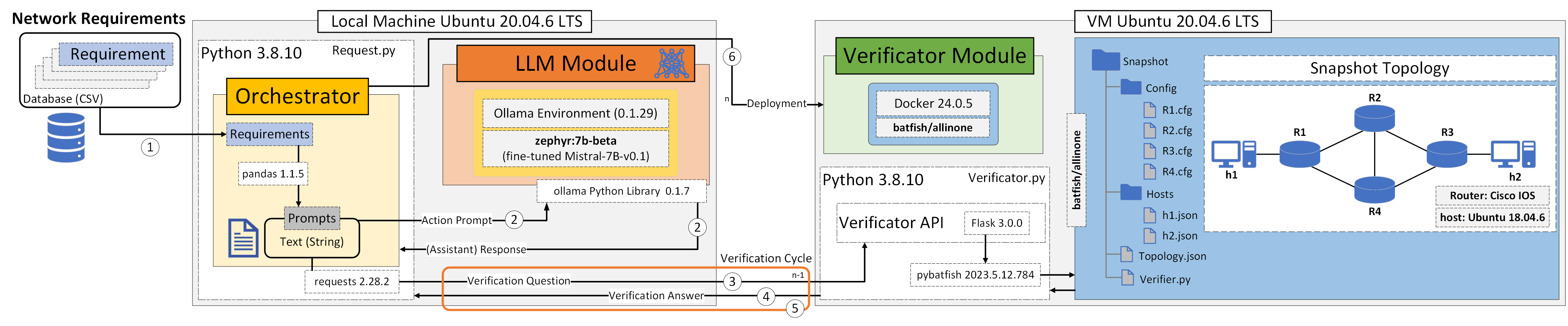}
\label{fig_V}
\end{figure*}

\subsection{LLM for ZSM Components}
LLMs can be used to carry out the ZSM components highlighted as follows.
\begin{itemize}
    \item \textit{Domain Analytics and E2E Analytics} services can be enriched with LLMs to provide accurate domain-specific recommendations and root cause analysis to support decision-making within a domain and services spanning multiple domains. LLMs can be used, for instance, to gain insights about consumer behavior and network status to create new intra and cross-domain service strategies. 
    \item LLMs can be integrated into \textit{Domain Intelligence and E2E Intelligence} services to accomplish closed-loop automation by offering decision assistance, decision making, and assistance in the plan of action using the knowledge obtained from Domain Analytics and E2E Analytics. LLMs can suggest, for example, adjustments to resource allocations or routing configurations to improve services' quality and efficiency. 
    \item In \textit{Domain Orchestration and E2E Orchestration} services, LLMs can support autonomous service deployment by generating plans and configurations based on intent requirements and constraints. LLMs can also evaluate the feasibility of novel services and provide recommendations for optimizing resource utilization. LLM-NetCFG can support autonomous service deployment by generating configurations based on intent requirements and network constraints. Additionally, LLM-NetCFG can evaluate the feasibility of changes previously to their deployment. 
    \item When involving cognitive and closed control loops, network management functional areas can be matched to self-healing (faults), self-configuration, self-monitoring (accounting), self-optimization (performance), and self-protection (security). Self-configuration can leverage an LLM to generate configurations from network status and device knowledge combined with external templates. Self-monitoring can integrate LLMs to analyze the data for identification purposes, such as anomaly detection based on flow data. Self-healing can use LLM decision-making to perform corrective actions when a failure is detected or predicted. LLMs can enrich self-optimization by creating a feedback from the outcomes of optimization interventions of the previous decision-making processes and learning from past experiences and outcomes to understand the system's behavior and performance characteristics. Self-protection can integrate specialized LLM learning agents to improve system security based on fine-tuning with well-known attack datasets.
    \item Intents are pivotal for achieving ZSM and, in general, autonomous networks; network management functions cannot evolve and mature to full closed-loop autonomy without intents. LLMs can be used to build up intent managers responsible for translating business intents to service intents, service intents to resource intents, and resource intents to network configurations. LLM-NetCFG uses intents to identify the administrator's goal and decompose it in the required configuration steps to be achieved before generation.
\end{itemize}

\subsection{Choosing LLMs For ZSM Components}
Selecting an LLM for each self-management functional area framed in ZSM involves mainly model size, which determines how well the model would understand and generate language; the number of tokens, which determines how well the model can interpret data; fine-tuning capability, which allows for the creation of specialized models tailored to specific tasks; and forecasting capability, which enables the model to make accurate predictions based on data.

\textbf{Self-configuration} demands an LLM large model size to manage the complexity and diversity of network configurations and a high token limit to process the detailed and extensive information involved. This self-property also needs fine-tuning support to adapt configurations to specific network environments and requirements.

\textbf{Self-monitoring} demands an LLM with a high token limit to process large volumes of flow data. This self-property also needs solid analytical and forecasting features to accurately identify issues or anomalies.

\textbf{Self-healing, self-optimization, and self-protection} need the forecasting capability to accomplish proactive operations. In addition, these self-properties require a large LLM model size to capture the complexity, and nuances of diverse network conditions and potential issues, as well as a high token limit to handle the detailed data involved in these tasks.

\section{LLM-NetCFG: Operation and Deployment}
This section introduces the proposed LLM-NetCFG to achieve network self-configuration using LLM.

\subsection{Overall operation}
The proposed  LLM-NetCFG comprises the modules LLM, Verifier, and Orchestrator, that allow an LLM model to generate and verify network configurations. The LLM module is responsible for the deployment and operation of an LLM in a local environment. It should guarantee that the model can receive questions and return answers. Also, it is required that the deployed model can be tuned so that model creativity can be nullified to receive consistent answers. The Verifier module receives a network configuration and checks its syntax errors. Moreover, it must check if the applied configuration achieves the primary goal of identifying inconsistencies or incongruities. The Orchestrator module coordinates automated tasks across the LLM module and Verifier module in different steps, allowing individual tasks to work together to generate and verify configurations.

LLM-NetCFG operates as follows (see Figure~\ref{fig_V}). \numrounded{1} The Orchestrator crafts custom prompts so the LLM module can achieve the classification and translation requirements expressed by the network administrator in natural language to low-level requirements. It also crafts other prompts so the LLM module can create the configurations; \numrounded{2} it uses the crafted prompts offered by the Orchestrator to perform the translation and configuration generation tasks (details in Section \ref{prompts}); before deployment \numrounded{3}, the Verifier module checks the correctness and applicability of every generated configuration (a configuration can include a set of individual configurations since meeting an intention can need modifying several devices set up); if this module detects errors or incongruities; \numrounded{4} the Verifier generates and sends a report with suggestions to improve the Orchestrator that, in turn, feedback (via a prompt) to the LLM-Generator with the actual configuration to make changes (tune the configuration from step \numrounded{2}); \numrounded{5} the Verifier notifies the Orchestrator when a configuration approves the verification process. A threshold value is necessary to avoid an infinite loop when the model does not generate a configuration that the Verifier approves; \numrounded{6} the LLM-Orchestrator sends the approved configuration for deployment to a ConfigsRepo.

LLM-NetCFG aims to provide the self-configure property to the ZSM architecture. It distinguishes itself from previous proposals by constructing a domain-adapted model that demands low resources; it has been specifically designed, trained, and optimized for self-configuration tasks.
An LLM model allows LLM-NetCFG to realize the ZSM configuration by converting a network intent described in a high-level natural language description into a configuration to achieve the network requirement goal. LLM-NetCFG integrates the validation step to prevent the model from introducing syntax errors when generating configuration commands. In particular, LLM-NetCFG is model-agnostic and modular, providing flexibility to integrate other management functions and facilitate its enhancement by specialized modules oriented to fine-tuning.

\begin{figure*}[!ht]
\centering
\caption{Classification, Translation and Generation Prompts - Example.}
\includegraphics[width=\textwidth]{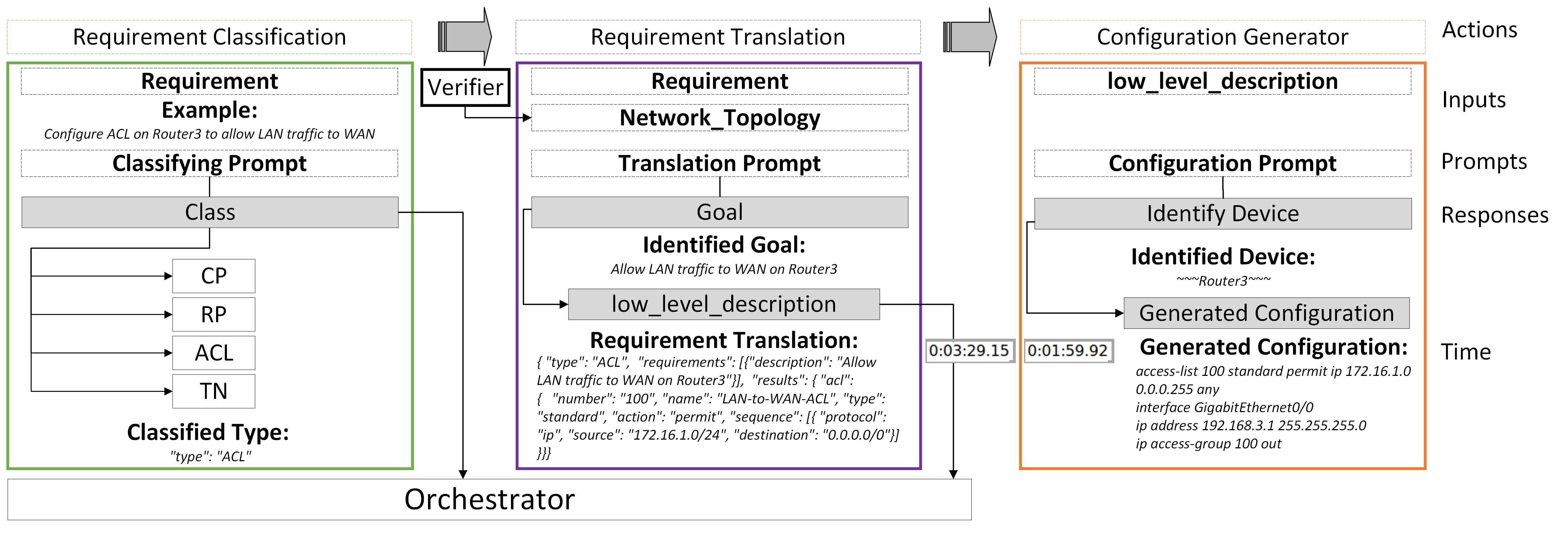}
\label{fig_Ex1}
\end{figure*}

\subsection{Prompts for Translation and Generation}
\label{prompts}

Figure \ref{fig_Ex1} exemplifies the LLM-Orchestrator tasks, automation prompts, and output examples. Initially, it uses the classification prompt to detect the configuration type received as input, an ACL described textually. Then, it uses the translation prompt to get a low-level description of such an input. Finally, it produces an executable configuration corresponding to the ACL rule by the configuration prompt.
Each prompt includes a \textit{system} role to instruct the LLM on the task to perform; 
an \textit{assistant} role to illustrate the LLM  output's shape, and a \textit{user} role to provide the LLM with the inputs for the task to perform. 

The \textit{system} of the Requirement Classification Prompt identifies the type of Intent. The Classification prompt indicates its \textit{assistant} role: \textit{``In the first line of your answer, specify \{type\}...''} of Intent. In the \textit{user} role, this prompt should input the Intent provided by the network administrator and the list of types of Intent. The \textit{type} can be, for instance, \textit{CP} for configuration interface properties.

The Requirement Translation Prompt specifies the
\textit{system} and \textit{assistant} roles as follows: \textit{``Your task is to take the high-level user \{Intent\} and define a \{low\_level\_description\} of the Intent in JavaScript Object Notation (JSON)}". In the \textit{user} role, this prompt must input the high-level requirement to translate and the status of the network to the LLM. The status can be obtained from the verifier. 

The Configuration Generation Prompt specifies the 
\textit{system} role as a network administrator responsible for creating network device configuration files to meet the translated Intent. The \textit{user} role provides input related to the general description of network devices and their brands and models. The \textit{assistant} role provides the answer structure that the LLM must generate. 

As LLMs usually explain their answers, it is necessary to inform them that they cannot provide explanations: \textit{``Just answer with the required configuration files to satisfy the \{low\_level\_description\}. You are not allowed to provide any explanation."}. Also, to assist Automation module in the verification process, a separator is used per device configuration: \textit{``Indicate at the beginning which device the configuration should be applied and separate each device section with the special characters \~~\~~\~~"}; this character is selected since it is not explicitly used in the configuration commands.

\section{Use Case: Interface, ACL, Router, and Tunnel Configuration}
This section introduces the proposed system by showing how it can generate and verify ZSM network configurations to support several tasks, such as creating and removing connections between devices and hosts, creating access control lists, setting OSPF routing rules, and creating secure communication using tunnel technologies.

\subsection{Prototype and Test Environment}
Figure~\ref{fig_V} presents the LLM-NetCFG prototype available at https://github.com/oscarGLira/LLM-based-Intelligent-Configuration-Validation-Framework.git. The Generator was developed using \textit{Zephyr 7B $\boldsymbol{\beta}$} \cite{zephyr}, an Open Source LLM fine-tuned by HuggingFace (i.e., a Mistral-based model), which excels in generating high-quality text and understanding complex prompts. Its improved context retention and fine-tuning performance make it adaptable to specific tasks. Its efficient processing and language generation suit tasks requiring intricate and well-defined responses.

The performance of Zephyr-7B has been evaluated using standardized benchmarks. A benchmark evaluated its general knowledge and reasoning capabilities in a wide range of topics (Abstract Reasoning Corpus). It was also tested for its ability to complete sentences contextually appropriately (HellaSwag) and its proficiency in university-level subjects such as humanities, science, and mathematics (Massive Multi-task Language Understanding). In addition, the model's accuracy was tested by providing accurate answers (TruthfulQA) or using common sense reasoning abilities (Winogrande).

Zephyr-7B was also evaluated using MT-Bench, which measures a model’s ability to handle conversations that follow specific instructions, and AlpacaEval, which assesses a model’s ability to respond to conversations based on one instruction. These benchmarks are used to rank LLMs. In these evaluations, Zephyr-7b scored 7.34 and 90.60, respectively, placing it at the top of all models of the same size. It outperformed other models, such as Falcon, Guanaco, Llama2, and Vicuna, with over 33 billion parameters, demonstrating the superior performance of Zephyr-7B. These findings show Zephyr-7B to be a reliable model. It performs better across various domains than larger models, especially in general knowledge, natural language understanding, and common sense reasoning, which strengthens the model's ability to perform the configuration generation task while being resource-efficient to be deployed locally with limited resources.

The Verifier was built as a Batfish \cite{batfish} application (i.e., an Open Source network analysis tool that allows the detection of errors and correctness in network configurations) using Pybatfish API and Flask (i.e., a Python library helpful in offering the Verificator as a REST-based service). 

The LLM-based Orchestrator and Generator modules were executed on a host machine with 16GB of RAM, a 2TB disk, and an Intel® Core™ i5 processor; this hardware configuration allowed \textit{Zephyr} to run efficiently. The Batfish application was executed on a Virtual Machine running Ubuntu 20.04 server with 4GB of RAM.

Figure~\ref{fig_V} also depicts the network for which the prototype generated configurations. The network follows a partial mesh topology with four Cisco routers. Two end-users are connected to R1 and R3. The topology and configuration details (about interfaces, OSPF for R1, R2, R3, and R4) are also available in a JSON file and in a configuration directory of our Git Hub.

\subsection{Experiments and Results}
In experimentation, we used a dataset with 90 resource configuration intents distributed in four types: CP, RP, ACL, and TN. Each Intent is a high-level requirement description of what the LLM-NetCFG is expected to achieve (Figure~\ref{fig_Ex1} exemplifies the Intent and Output configurations). CP, RP, ACL, and TN include requirements related to interfaces, access lists, routing, and tunnel configurations, respectively, and are required in the verification process.

\begin{figure}[!ht]
\centering
\caption{Accuracy in classifying requirements' type}
\includegraphics[width=0.35\textwidth]{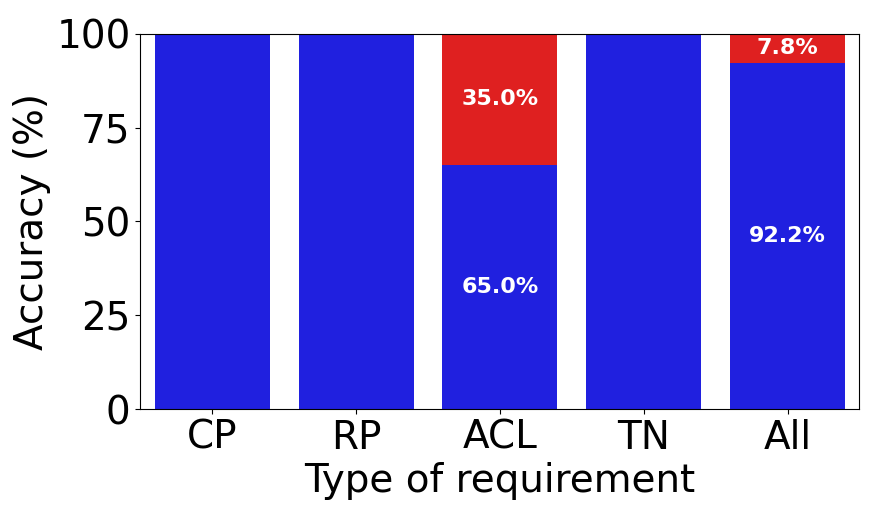}
\label{fig_Types}
\end{figure}

Figure \ref{fig_Types} shows how LLM-NetCFG correctly classified $92.2\%$ of requirements, which means that it efficiently understands the intents' goals, gathers relevant configuration information (e.g., from the Internetwork Operating System commands), and generates specific goals (in the tests for Cisco devices). LLM-NetCFG created a new class named $Other$ for misclassified requirements. It had problems detecting not well-described ACL requirements; for example, an ACL defined to filter traffic SNMP was classified as $Other$ (i.e., SNMP, a non-defined class) and not as ACL. The $Other$ class highlights the LLM's hallucination problem, which can negatively impact ZSM configurations. It underscores the importance of precise high-level requirement definitions for optimal model understanding. Misclassification can affect the verification process, as the Orchestrator may only perform a syntax check rather than a specific verification tailored to the requirements, posing a significant challenge for integrating LLMs into real networking solutions.

\begin{figure}[!ht]
\centering
\caption{Total Processing Time By Requirement Complexity For Each Class.}
\includegraphics[width=0.35\textwidth]{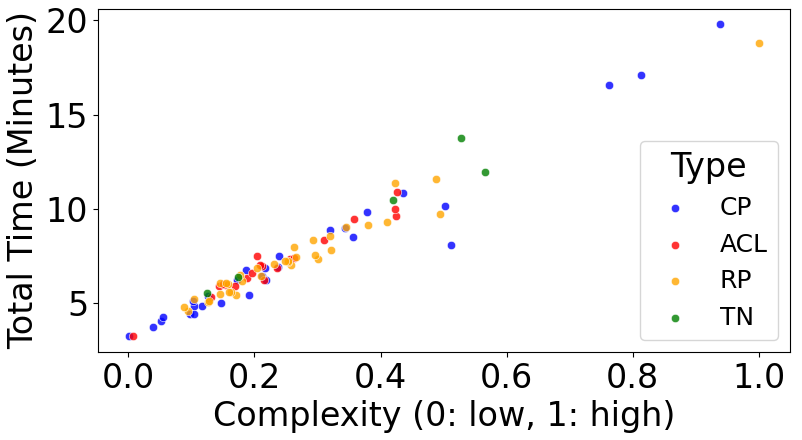}
\label{fig_Times}
\end{figure}

Figure~\ref{fig_Times} shows that LLM-NetCFG spent five to seven minutes handling non-complex intent goals (incorporating a new device in the network, turning off a network interface, and setting up an access control list). This result is because the system needed just a cycle of classification and translation of the intent's goal and generation and verification of configuration to meet it. The LLM-NetCFG took more than 10 minutes to cope with complex intents (including several goals) because it needed several verification cycles. Using fine-tuning techniques to reduce the model's time response for facing tasks with time requirements is also a relevant challenge to beat when realizing LLM-based network management.

\section{Future Research Directions}
\subsection{LLMs for ZSM Management Functions: What can be done?}

\textbf{LLM for Self-configuration.} Device configuration troubleshooting extends beyond merely verifying syntax errors. Even when configurations appear error-free, they can still be insecure due to inadequate practices. LLMs can be integrated as sophisticated configuration correctness checkers, ensuring compliance and security throughout the configuration process. As LLM-based systems have improved coding practices, they can enhance automatic configuration systems.

\textbf{Intent-handling with LLM.} Translating business goals into technical configurations can be complex and error-prone, impacting operational efficiency. An LLM-based autonomous intent mapper that interprets natural language input can streamline self-configuration by converting business goals into precise technical settings. The intent mapper can enhance operational efficiency and reduce configuration errors.

\textbf{LLM for Self-healing.} Current ML-based methods for failure localization need to be scalable to analyze high-dimensional logs in moderate-size networks. Integrating LLMs into the ZSM architecture can assist in fault detection, localization, and diagnosis. When jointly used monitoring tools, LLMs can make the system more proactive in self-healing. Efficient data analysis techniques for leveraging LLMs' Knowledge Augmentation capabilities should be incorporated into network management systems  capable of self-healing.

\textbf{LLM for Self-monitoring.} Managing and analyzing large volumes of network data to generate actionable insights can be challenging. LLMs can overcome such a challenge with a zero-touch data analytics capability that enhances decision-making, improves service performance, and facilitates proactive network optimization by providing real-time visibility into network performance and usage patterns. Also, ZSM Data Services can benefit from LLMs for Root Cause Analysis, making identifying and resolving network issues more manageable and timely.

\textbf{LLM for Self-optimization.} Due to the large volumes of data involved, detecting traffic patterns that lead to network performance degradation is complex. LLMs could assist in self-optimization by analyzing extensive data to detect the network segments, links, and devices that avoid QoS and QoE meetings, which is pivotal for achieving more reliable and efficient network performance.

\textbf{LLM for Self-securing.} Detecting anomalies and security threats in large networks can be complicated. LLMs can enhance anomaly detection systems by supporting the understanding and processing extensive network logs, alerts, and contextual information to identify deviations from normal behavior. Also, LLMs can help security teams uncover the underlying causes of security incidents and breaches and generate actionable recommendations for mitigating risks, which is fundamental to self-healing.

\subsection{Challenges Using LLMs}
When integrating LLMs into the ZSM architecture, it is essential to note that these models present specific intrinsic challenges to their architecture and characteristics that need to be addressed. The most relevant are discussed below.  

\textbf{Hallucinations.} LLMs may generate incorrect, nonsensical, or fake responses. Techniques like fine-tuning and prompt engineering, using refined and domain-specific information, can mitigate these hallucinations and improve performance. 

\textbf{Explainable LLMs.} LLMs work as opaque models, which may limit their adoption in network management. Investigating explainability techniques, such as using the same model to justify its actions, can provide interpretability and transparency to LLM-based solutions.

\textbf{LLM for Networks.} Since there is no "Theory of LLM for Networks" for realizing network management with LLMs, specific datasets must be worked around to leverage LLMs’ potential in networking. Additionally, coordinating LLM-based FCAPS into a closed cognitive control loop requires an orchestration layer to ensure that intents are met at the business, service, and resource levels.

\section{Conclusion}
Using LLMs to realize intent-driven self-capabilities framed into ZSM opens up a wide range of research directions and application possibilities for zero-touch networks. Leveraging the full potential of LLMs in autonomous network management will require devising novel frameworks for fast prototyping and proper training, fine-tuning and orchestrating FCAPS, and adequate interfacing between large models and real networks to ensure effectiveness and minimize exposure to security incidents. This paper presented how LLMs can be used to design ZSM self-configuration agents, introduced LLM-NetCFG to automatically configure network devices from intents, and delved into the opportunities and challenges of integrating LLMs into ZSM.

\bibliographystyle{IEEEtran}
\bibliography{references}
\vspace{-1.3cm}
\begin{IEEEbiographynophoto}
{M.Sc Oscar G. Lira} (GS'17) is pursuing his Ph.D. degree in Computer Science at the State University of Campinas, Campinas, Brazil.
\end{IEEEbiographynophoto}
\vspace{-1.3cm}
\begin{IEEEbiographynophoto}
{Ph.D. Oscar M. Caicedo}(GS'11--M'15--SM'20) is a full professor at the Universidad del Cauca, Colombia.
\end{IEEEbiographynophoto}
\vspace{-1.3cm}
\begin{IEEEbiographynophoto}
{Ph.D. Nelson L. S. da Fonseca} (M'88--SM'01) 
is a Full professor at the Institute of Computing, State University of Campinas, Campinas, Brazil. 
\end{IEEEbiographynophoto}
\end{document}